\documentclass[conference]{IEEEtran}
\IEEEoverridecommandlockouts
\usepackage{cite}
\usepackage{amsmath,amssymb,amsfonts}
\usepackage{algorithmic}
\usepackage{graphicx}
\usepackage{textcomp}
\usepackage{xcolor}
\usepackage{float}
\usepackage{setspace}
\usepackage{tikz}
\usepackage{algorithm}
\usepackage{algorithmic}
\usepackage{bm}
\usepackage{subfigure}
\usepackage[utf8]{inputenc}
\input{arraymac.tex}

\def\BibTeX{{\rm B\kern-.05em{\sc i\kern-.025em b}\kern-.08em
    T\kern-.1667em\lower.7ex\hbox{E}\kern-.125emX}}

\begin{document}

\title{RIS-Assisted Interference Mitigation \\ for Uplink NOMA\\
{\footnotesize \textsuperscript{}}
\thanks
}

\author{\IEEEauthorblockN{Azadeh Tabeshnezhad}
\IEEEauthorblockA{\textit{Electrical Engineering} \\
\textit{Chalmers University of Technology}\\
Gothenburg, Sweden \\
azadeh.tabeshnezhad@chalmers.se}
\and
\IEEEauthorblockN{A. Lee Swindlehurst}
\IEEEauthorblockA{\textit{Electrical Engineering and Computer Science} \\
\textit{University of California, Irvine}\\
Irvine, USA \\
swindle@uci.edu}
\and
\IEEEauthorblockN{Tommy Svensson}
\IEEEauthorblockA{\textit{Electrical Engineering} \\
\textit{Chalmers University of Technology}\\
Gothenburg, Sweden\\
tommy.svensson@chalmers.se }
\and
}

\maketitle

\begin{abstract}
 Non-orthogonal multiple access (NOMA) has become a promising technology for next-generation wireless communications systems due to its capability to provide access for multiple users on the same resource. In this paper, we consider an uplink power-domain NOMA system aided by a reconfigurable intelligent surface (RIS) in the presence of a jammer that aims to maximize its interference on the base station (BS) uplink receiver. We consider two kinds of RISs, a regular RIS whose elements can only change the phase of the incoming wave, and an RIS whose elements can also attenuate the incoming wave. Our aim is to minimize the total power transmitted by the user terminals under quality-of-service constraints by controlling both the propagation from the users and the jammer to the BS with the help of the RIS. The resulting objective function and constraints are both non-linear and non-convex, so we address this problem using numerical optimization. Our numerical results show that the RIS can help to dramatically reduce the per user required transmit power in an interference-limited scenario. 
\end{abstract}
\vspace{.3cm}
\begin{IEEEkeywords}
NOMA, RIS, Phase shift, Optimization
\end{IEEEkeywords}

\section{Introduction}
Recently, the rapid development of the internet of things (IoT), multimedia applications, and vehicle-to-everything (V2X) communications have led to an increase in the number of user equipment (UE) and the need for enhanced data rates for 5G-and-beyond mobile wireless networks \cite{b1}. To address these demands, some potential candidate technologies have been proposed including massive multiple-input multiple-output (MIMO) antenna arrays \cite{b2}, ultra-dense networks \cite{b3}, millimeter wave communications \cite{b4}, and non-orthogonal multiple access (NOMA) \cite{b5}. All of these techniques are useful for enabling massive connectivity among many wireless devices. NOMA is of particular interest because it allows multiple users to share the same orthogonal time-, frequency-, spatial- and code-domain resource blocks \cite{b6}. In \cite{b7} NOMA was shown to give an improvement in system throughput and user-fairness for a single-input single-output (SISO) channel compared to OFDMA. Since then, many researchers have investigated the benefits of NOMA for next-generation radio access techniques. However, there is substantially less work on uplink NOMA, and there is a lack of work addressing the reliability of NOMA in the presence of an active jammer.

On the other hand, the use of RIS has emerged as a unique technology for improving both spectral and energy efficiency. An RIS consists of an array of elements whose reflective properties can be individually controlled, enabling some degree of control of the wireless propagation environment \cite{b9}. In a conventional RIS implementation, it is the phase shift of the reflection coefficient of each element that is adjusted in order to achieve the desired effect on the wireless channels. More recently, researchers have also studied RIS architectures where both the phase shift and the attenuation of the reflection coefficient of each element can be individually controlled, so-called absorptive RIS (A-RIS). In this latter case, the energy absorbed by the A-RIS can be refracted to directions on the other side of the surface \cite{b16},\cite{b17} or sampled using an active radio frequency (RF) receiver for channel estimation or sensing \cite{b18},\cite{b20}. RIS technology has been applied in many different types of wireless communications scenarios, including NOMA \cite{b8}. However, while the use of conventional phase-shift-only RIS has been proposed for NOMA applications with the goal of improving spectral efficiency, to our knowledge there is no work reported on using A-RIS with NOMA, nor on using RIS to mitigate the impact of external interference (e.g. jamming) on NOMA performance.  

In this paper, we address these two avenues and show how a regular RIS and an A-RIS can provide improved rejection of a multi-antenna jammer on a simple two-user uplink NOMA system. The paper outline is as follows. In Section \ref{sec:Overview} we deepen the analysis of related works on NOMA and RIS. In Section \ref{sec:SystemModel} we detail our targeted scenario, provide our RIS-assisted NOMA system model, and define our targeted optimization problem. In Section \ref{sec:Results} we present our obtained results using numerical optimization on RIS-assisted jammer interference mitigation in uplink NOMA, and we summarize our findings in Section \ref{sec:Conclusions}.

\section{Overview of NOMA and RIS}
\label{sec:Overview}
\subsection{Related work}
\begin{figure}
    \centering
    \includegraphics[height=3.3cm]{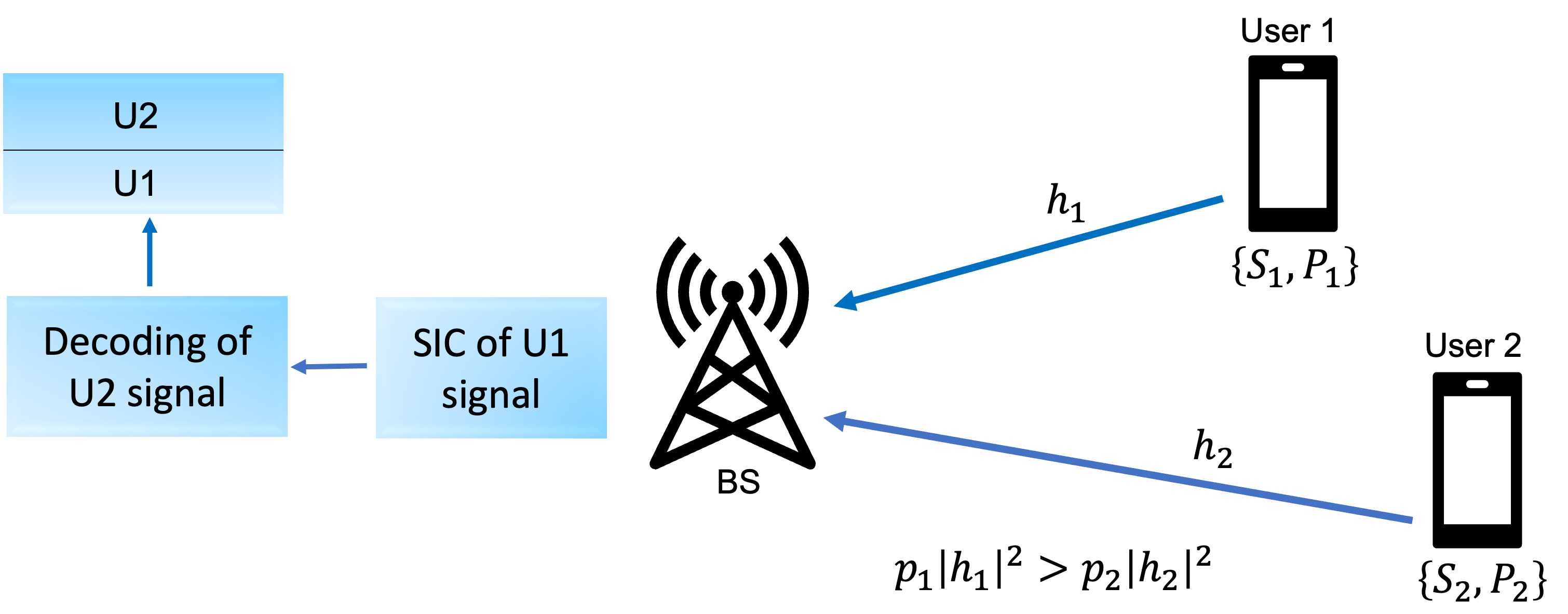}
    \caption{Illustration of uplink power-domain NOMA with two users sending message $S_1$, and $S_2$, with power 
    $p_1 \leq P_1$ and $p_2 \leq P_2$ respectively.}
    \label{fig1}
\end{figure}
\textit{(1) Literature on NOMA}: NOMA is an interesting technique since it enables multiple access for users in the same resource. Generally, NOMA is divided into two main classes: power-domain and code-domain \cite{b10}. Power-domain NOMA exploits situations where the users have different path loss levels. In downlink NOMA this implies that the users nearer to the base station (BS) typically have better channel conditions compared to distant users, who require higher transmission power to mitigate the higher path loss. The idea behind power-domain NOMA is that the users nearer the BS can employ successive interference cancellation (SIC) to remove the strong signal destined for the remote users before decoding their own signal \cite{b11}. In uplink NOMA, which is the focus of this paper, the BS can similarly employ SIC for the nearer users to remove those before decoding the signals for the remote users. However, as illustrated in Fig.~\ref{fig1} for the case of two users, the SIC condition needs to include both the individual users' transmit power $p_i$, as well as the channel gain $| h_i |$ for paired users $i=1, 2$ in the same resource.

The superiority of NOMA over conventional OMA is derived from the improved spectral efficiency due to the sharing of time/frequency/space/code resources, its allocation of different quality-of-service (QoS) levels to users based on power level allocation, and the fact that it supports massive connectivity, lower latency, and an enhanced cell-edge user experience \cite{b12}. NOMA technology has been combined with other technologies such as massive MIMO, cognitive-communication, integrated access and backhaul (IAB), and RIS.

\textit{(2) Literature on RIS}: An RIS is a surface consisting of electromagnetic material. The surface usually consists of a large number of low-cost reflecting elements whose properties can be controlled. In a conventional RIS design, the phase of the reflection coefficient of each element is controlled in order to enable modification of the propagated waves \cite{b13}. More recently, RIS designs have been proposed that enable control of both the phase and the attenuation of each element, providing additional degrees of freedom for shaping the resulting wave field \cite{b18},\cite{b21}. RIS technology has been proposed for many different types of wireless communication scenarios, including NOMA \cite{b8},\cite{b13}. It has been shown that the performance of NOMA systems can be improved by increasing the number of reflecting elements \cite{b14}, and the spectral efficiency can be significantly enhanced with RIS-assisted NOMA, compared to NOMA without RIS and traditional OMA \cite{b8}.


\section{RIS-NOMA Network: System Model}
\label{sec:SystemModel}
In this section, we first describe the system and signal model for our considered RIS-NOMA  system.

\subsection{System Model}\label{AA}
\begin{figure}
    \centering
    \includegraphics[height=6cm]{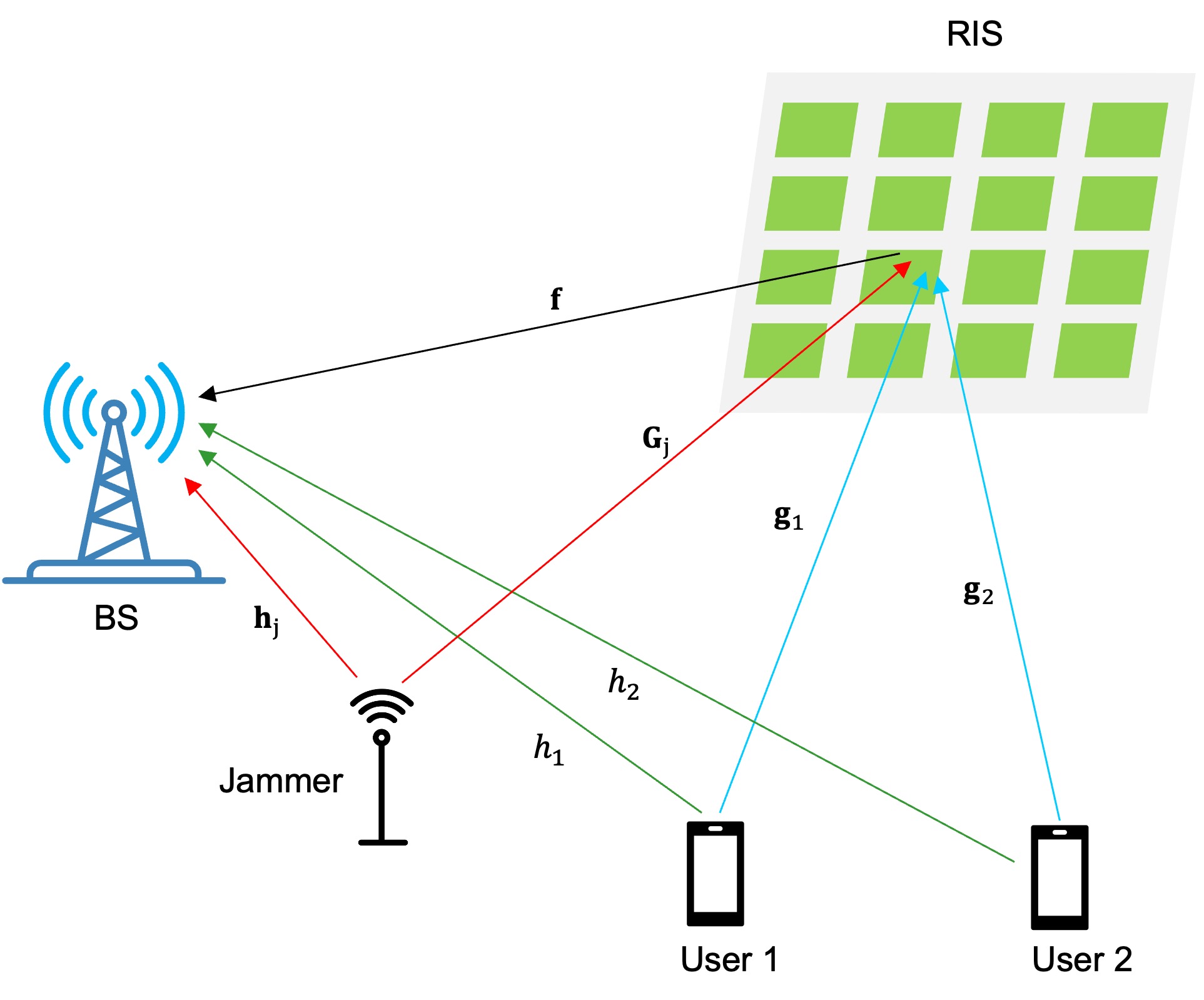}
    \caption{Illustration of RIS-NOMA aided two users uplink communication.}
    \label{fig2}
\end{figure}
We consider an uplink transmission RIS-NOMA system where the BS is equipped with a single-antenna, an RIS with $N$ elements, a jammer with $M$ antennas, and two single antenna users using PD-NOMA, as shown in Fig.~\ref{fig2}. The received signal at the BS can be written as
\begin{equation} \label{eu_1}
r = \sum_{i=1}^2 \left( h_i + \mathbf{f}^T \mathbf{\Phi}\mathbf{g}_i \right) x_i + \left( \hbf_\text{j}^T + \mathbf{f}^T \mathbf{\Phi}\Gbf_\text{j}\right) \xbf_\text{j} +n_\text{r},
\end{equation}
where $h_i \in \mathbb{C}^{1\times 1}$ denotes the direct channels between the users and the BS, $x_i$ is the symbol transmitted by user $i$, $\mathbf{f} \in \mathbb{C}^{N\times 1}$ denotes the channel vector between the RIS and the BS, $\mathbf{\Phi} = \text{diag} \{ e^{j\theta_1},  e^{j\theta_2},..., e^{j\theta_N}\}\in \mathbb{C}^{N\times N}$ is the diagonal matrix containing the RIS phase-shifts, $\theta_n\in [0,2\pi]$ is the phase shift of the $n$-th reflecting element, $\bm {g}_i \in \mathbb{C}^{N\times 1}$ denotes the channel vector between user $i$ and the RIS, $\bm {h}_\text{j} \in \mathbb{C}^{M\times 1}$ denotes the channel vector between the jammer and the BS,  $\mathbf{G}_\text{j} \in \mathbb{C}^{N\times M}$ denotes the channel matrix between the jammer and RIS, $\xbf_\text{j} \in \mathbb{C}^{M\times 1}$ is the signal transmitted by the jammer, and $n_\text{r}$ $\sim$ $\mathcal{CN}$ $(0,\sigma^2)$ is additive white Gaussian noise (AWGN). We define the transmit power of each user as $p_i = E(|x_i|^2)$.

Assuming that UE$_1$ has the strongest channel and using a sufficiently large transmit power level $p_1$ fulfilling the condition in Fig.~\ref{fig1}, $p_1 | h_1 |^2 > p_2 | h_2 |^2$, the BS 
would first decode the signal for UE$_1$ treating the interference by the signal from UE$_2$ as AWGN, and then subtracting it from the received signal $r$ when decoding the signal from UE$_2$. Thus, the signal to interference plus noise ratio (SINR) of UE$_1$ is given by
\begin{equation} \label{eu_2}
    \gamma_1 = \frac{p_1\vert h_1 + \mathbf{f}^T \mathbf{\Phi}\mathbf{g}_1 \vert^2 }{p_2\vert  h_2 +\mathbf{f}^T\mathbf{\Phi} \mathbf{g}_2 \vert^2 + \sigma_\text{j}^2 + \sigma^2} \; ,
\end{equation}
where we let $\sigma_\text{j}^2$ denote the power of the jammer signal received by the BS. This term will be derived in Subsection \ref{sec:Jammer}. Under the assumption of perfect SIC of the UE$_1$ signal, i.e. no error propagation, the SINR for UE$_2$ can be written as
\begin{equation} \label{eu_3}
   \gamma_2 = \frac{p_2\vert  h_2 +\mathbf{f}^T\mathbf{\Phi} \mathbf{g}_2 \vert^2 }{\sigma_\text{j}^2 +\sigma^2} \; .
\end{equation}

\subsection{Absorptive RIS}
As mentioned above, recent work has considered RIS implementations in which not all energy is reflected. In prior work, the non-reflected energy is either transmitted or ``refracted'' to the other side of the RIS, or it is demodulated and sampled for channel estimation or sensing purposes. Here we simply assume that the RIS absorbs an adjustable fraction of the incoming energy at each element, without assuming that the absorbed energy is used for any other purpose. We will see that the ability to adjust the magnitude of the reflected power at each RIS element provides additional degrees of freedom (DoFs) that are particularly useful in situations like the one we consider in this paper where interference mitigation is required. Mathematically, the A-RIS model assumes that the reflection coefficient of each RIS element can be described as $\beta_n e^{j\theta_n}$, where $0 \le \beta_n \le 1$ describes the amplitude of the reflected signal component. Such a constraint has numerical advantages as it is convex. In practice, the value of $\beta_i$, like $\theta_n$, may have to be quantized, and the two variables are very likely to be coupled, meaning that changes to one will affect the other. For this initial study, we ignore such effects in order to investigate the potential gains in an idealized scenario.

\subsection{Smart Jammer Modeling}
\label{sec:Jammer}
We assume a ``smart'' jammer that is aware of $\Phibf$ and the channel state information (CSI) of all the channels used by the jammer, i.e. \{$\mathbf{h}_\text{j}$, $\mathbf{f}$, $\mathbf{G}_\text{j}$\}, and whose goal is to choose $\xbf_\text{j}$ such that the interference power at the BS is maximized.  In other words, the jammer designs $\xbf_\text{j}$ based on the following problem:

\begin{equation} \label{eu_4}
\max_{\xbf_\text{j}} \; \| \left(\hbf_\text{j}^T + \fbf^T\Phibf\Gbf_\text{j}\right) \xbf_\text{j} \| \quad \mbox{\rm s.t.} \quad E \left(\|\xbf_\text{j}\|^2\right) \le P_\text{j} \; ,
\end{equation}
where $P_\text{j}$ is the per antenna jammer transmit power. The solution to this problem is simply $\xbf_\text{j}=(\sqrt P_\text{j}/\rho) \left(\hbf_\text{j}^T+\fbf^T\Phibf\Gbf_\text{j}\right)^H$, where $\rho=\|\hbf_\text{j}^T+\fbf^T\Phibf\Gbf_\text{j}\|$. Thus, the term due to jamming in the denominator of the SINR expressions (\ref{eu_2}) and (\ref{eu_3}) becomes:
\begin{equation} \label{eu_5}
 \sigma_\text{j}^2 = P_\text{j}\| \hbf_\text{j}^T+\fbf^T\Phibf\Gbf_\text{j} \|^2 \; .
\end{equation}

\subsection{Optimization Problem}
There are a number of ways to illustrate the benefit of the A-RIS in the problem under consideration. Here we study the problem of minimizing the total transmit power of the users such that given SINR quality-of-service constraints of the users are met. The problem can be formulated mathematically as follows.
\begin{equation}\label{eu_6} 
\begin{aligned}
     &\min \quad p_1 + p_2 \quad \textrm{such that}\\                   
             & \mathbf{C1}: p_1 \geq 0 \\
             & \mathbf{C2}: p_2 \geq 0\\
             & \mathbf{C3}: \frac{p_1\vert  h_1 +\mathbf{f}^T \mathbf{\Phi} \mathbf{g}_1 \vert^2 }{P_\text{j}\| \mathbf{h}_\text{j}^T + \mathbf{f}^T \mathbf{\Phi} \mathbf{G}_\text{j}\|^2 +p_2\vert  h_2 + \mathbf{f}^T \mathbf{\Phi} \mathbf{g}_2 \vert^2 + \sigma^2} \geq T_1\\
            & \mathbf{C4}: \frac{p_2\vert  h_2 +\mathbf{f}^T \mathbf{\Phi} \mathbf{g}_2  \vert^2 }{P_\text{j}\|\mathbf{h}^T_\text{j} +\mathbf{f}^T \mathbf{\Phi} \mathbf{G}_\text{j} \|^2+\sigma^2} \geq T_2\\
            &\mathbf{C5}:0\leq\beta_n\leq 1, \; n=1,\cdots,N\\
            &\mathbf{C6}:0\leq\theta_n\leq 2\pi, \; n=1,\cdots,N\\
\end{aligned}
\end{equation}

Constraints {\bf C1} and {\bf C2} ensure that the transmitted power is non-negative, {\bf C3} and {\bf C4} correspond to the desired SINR constraints for user 1 ($T_1$) and user 2 ($T_2$), respectively, and {\bf C5} and {\bf C6} enforce the properties of the A-RIS. When we consider the performance of a conventional phase-only RIS, constraint ${\bf C5}$ is replaced with $\beta_n=1$. 
As seen in (\ref{eu_4}), the resulting objective function and constraints are both non-linear and non-convex, so we address this problem using numerical optimization.

\section{Simulation Results}\label{sec:Results}
In this section, we study the relative performance of the considered system with an A-RIS, with a conventional phase-only RIS, and without any kind of RIS. 

We assume that all channels in our system model illustrated in Fig.~\ref{fig2} are Rayleigh fading, but with different channel gains, and we solve the optimization problem (\ref{eu_6}) using the parameters in Table 1. In the results, we present the required total power for the two UEs to meet the target SINRs $T_1$ and $T_2$ normalized with the BS noise, i.e. $(p_1+p_2)/\sigma^2$. In all the presented results, the resulting SINRs $\gamma_1$ and $\gamma_2$ for UE$_1$ and UE$_2$, respectively, are verified to satisfy the targets $T_1$ and $T_2$, respectively. In fact, $T_1$ and $T_2$ are met with equality, which indicates the optimality of the numerical optimization results.

 \begin{table} [htb]
  \label{table1} 
    \caption{Summary of key parameters}
    \begin{tabular}{|l|l|l|}
    \hline
    \bf{Variable}    &  \bf{Description}            & \bf{Value} \\ \hline 
    $\sigma^2$       &  AWGN power at BS receiver   & 1 \\ \hline
    $P_\text{j}/\sigma^2$     &  Normalized jammer antenna power  & 40   \\ \hline 
    $\mathbb{E} \{ h_1 \}$    &  Gain between UE$_1$ \&  BS    & 5 \\ \hline 
    $\mathbb{E}\{ h_2 \}$        &  Gain between UE$_2$ \&  BS    & 2   \\ \hline 
    $\mathbb{E}\{\mathbf h_\text{j}(:)\}$   &  Gain between jammer   \&  RIS    & 1     \\ \hline 
    $\mathbb{E}\{\mathbf{g_1(:)}\}$   &  Gain between UE$_1$ \& RIS    & 1    \\ \hline 
    $\mathbb{E}\{\mathbf{g_2(:)}\}$   &  Gain between UE$_2$ \&  RIS   & 0.2    \\ \hline 
    $\mathbb{E}\{\mathbf{G_\text{j}(:,:)}\}$   &  Gain between jammer \&  RIS    & 0.2  \\ \hline 
    $\mathbb{E}\{\mathbf{f(:)}\}$     &  Gain between RIS \&  BS      & 1       \\ \hline 
    $T_1$            &  Targeted SINR threshold for UE$_1$  & 5\\ \hline 
    $T_2$            &  Targeted SINR threshold for UE$_2$  & 5 \\ \hline 
    \end{tabular}
\end{table}

\begin{figure*}
    \centering
    \subfigure[$M=4$]{\includegraphics[width=.329\textwidth]{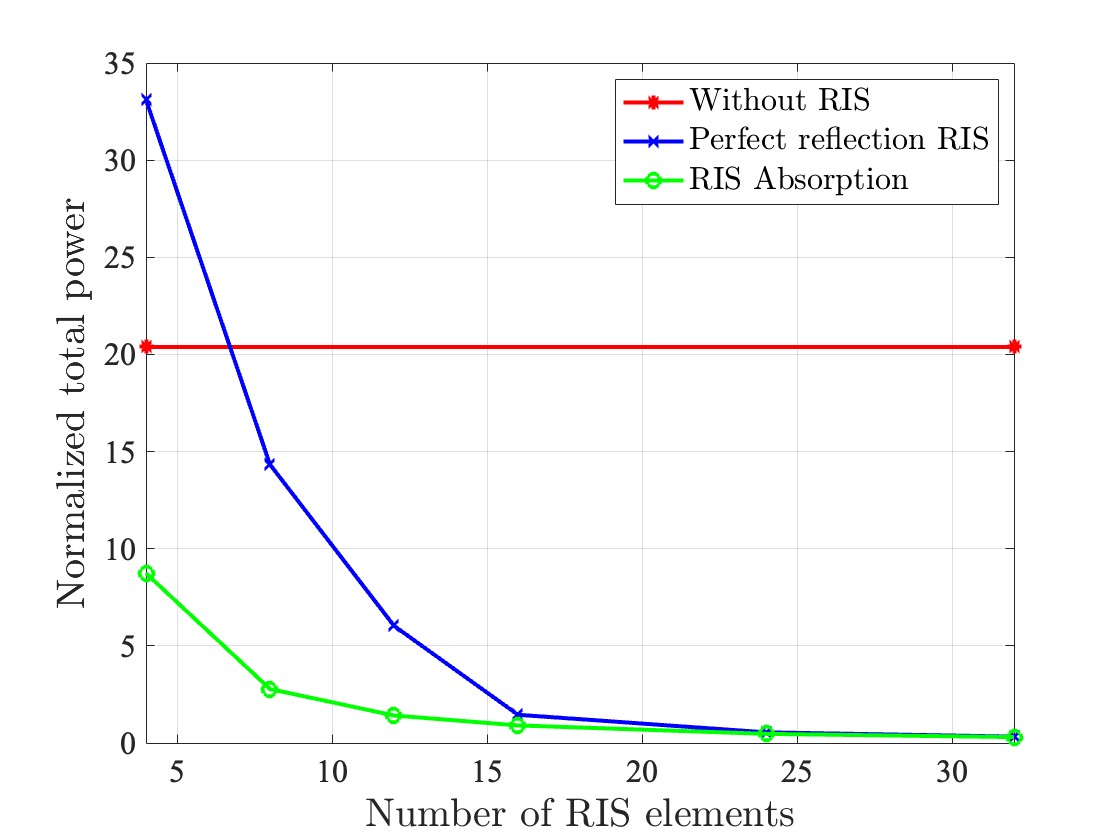}\label{fig3}}
    \subfigure[$M=8$]{\includegraphics[width=.329\textwidth]{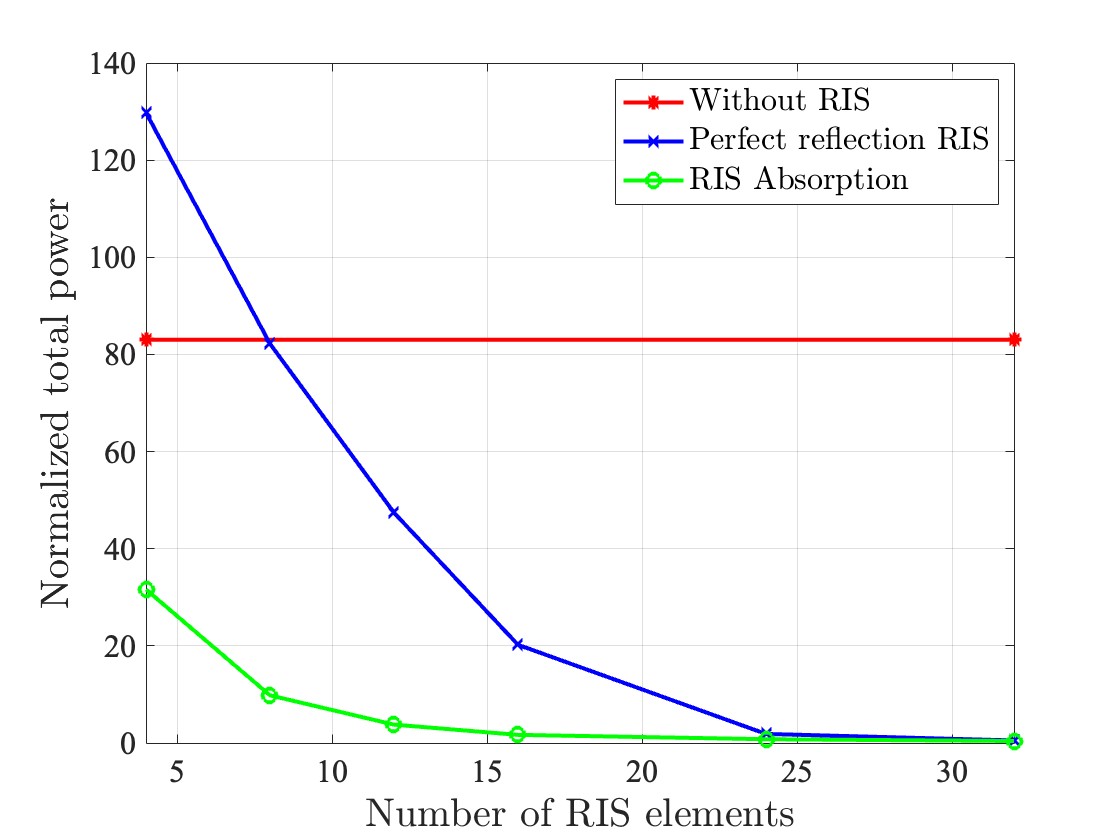}\label{fig4}} 
     \subfigure[$M=16$]{\includegraphics[width=.329\textwidth]{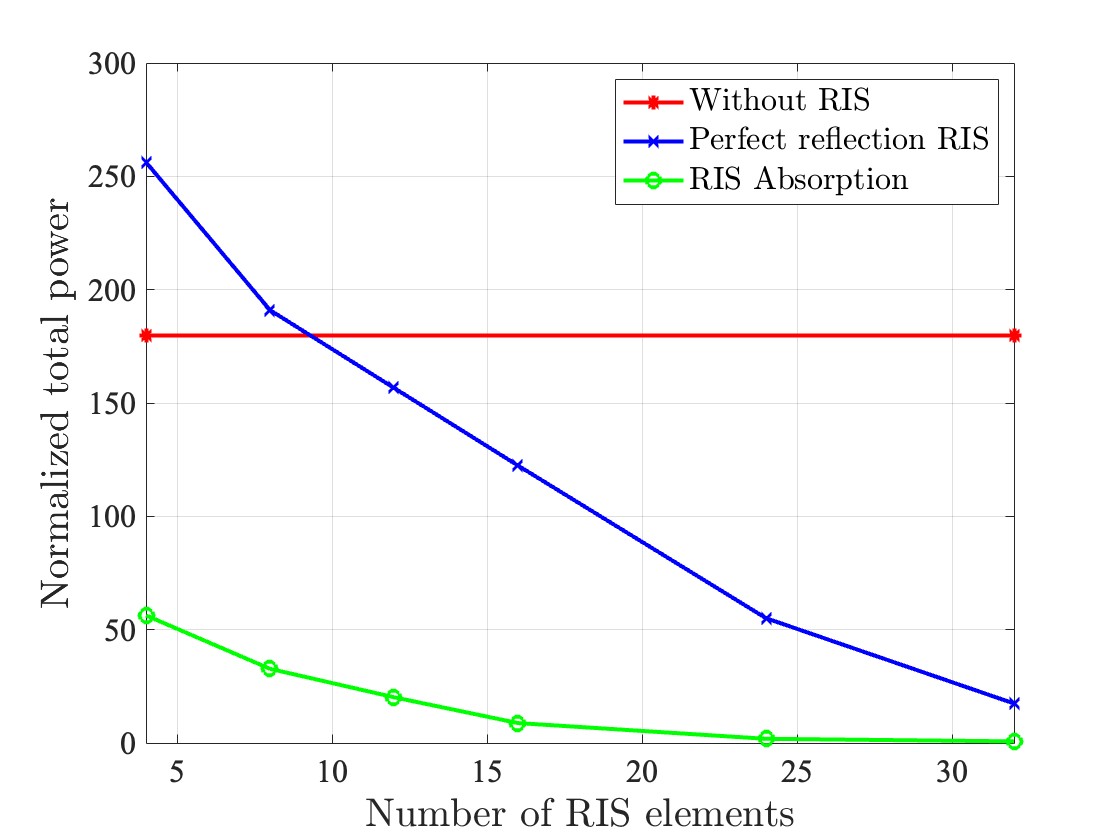}\label{fig5}}   
     \caption{Required total transmit power to meet the user quality of service requirements in the presence of a jammer with $M=4, 8, 16$ antennas, as a function of the number of RIS elements.}
    \vspace{.2cm}
\end{figure*}

In Fig.~\ref{fig3}, we show the resulting required total transmit power as a function of the number of RIS elements in the presence of a jammer with $M=4$ antennas. As can be seen, a regular non-absorbing RIS is beneficial compared with no RIS provided the RIS is equipped with more than $N=4$ elements, whereas the A-RIS has a performance gain for all considered $N$.

In Fig.~\ref{fig4} and Fig.~\ref{fig5}, we show the corresponding results for a jammer with $M=8$ and $M=16$ elements, respectively. The relative performance for the no RIS, regular RIS, and A-RIS is the same, but due to the stronger jammer, the crossing point for the regular RIS increases with the number of jammer antennas $M$. It is also interesting to note that, compared to the no RIS case, the required total transmit power more than doubles when the number of jammer antennas is increased from $M=4$ to $M=8$, although the total jammer power is only doubled. That can likely be explained by the additional beamforming gain for the jammer. Increasing the number of jammer antennas from $M=8$ to $M=16$, however, results in just slightly more than doubled required total transmit power, which can most likely be explained by saturation in the beamforming gain. Another interesting observation is that when the RIS is very small, e.g., with only 4 or 8 elements, the performance with a conventional phase-only RIS is actually worse than if no RIS were present at all. This is because the conventional RIS does not have a sufficient number of degrees of freedom to eliminate the interference, and its presence as a reflector only serves to actually increase the interference seen by the BS. The A-RIS is able to control the amount of reflected energy in addition to the phase shifts and thus does not suffer from this drawback.

\begin{figure}
    \centering
    \includegraphics[height=7cm]{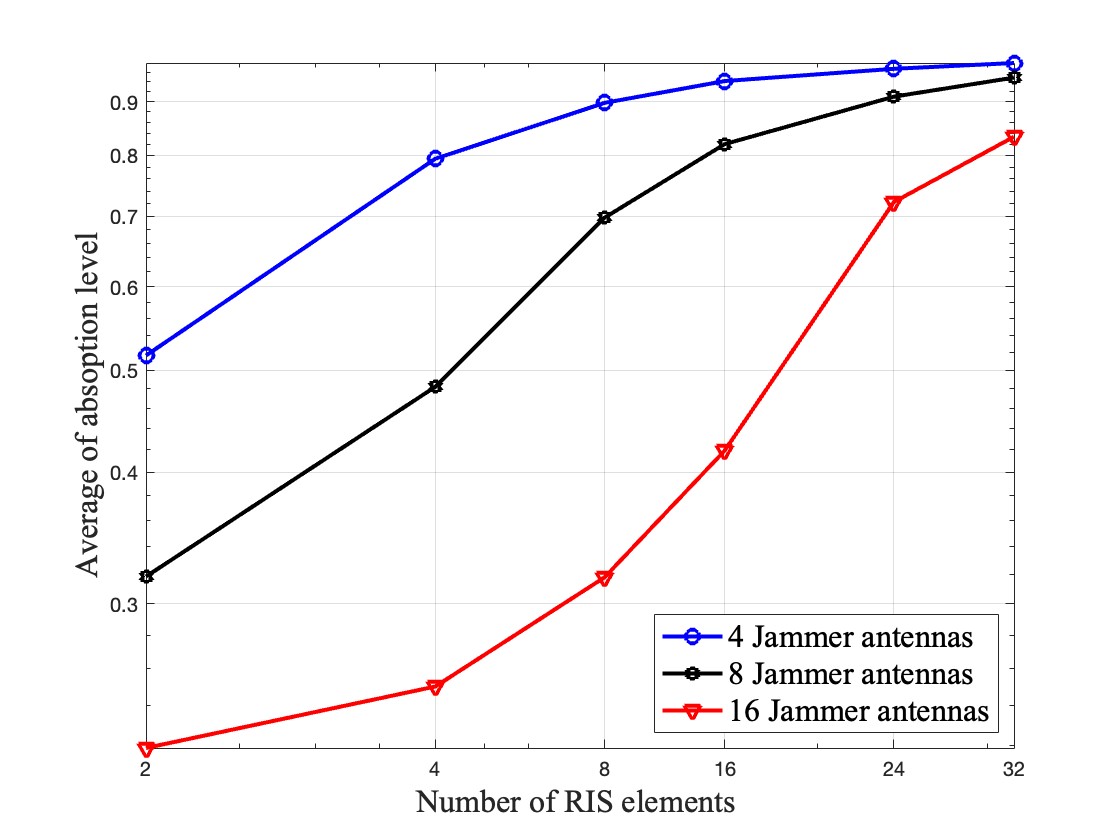}
    \caption{Resulting average absorption level of the A-RIS as a function of the number of A-RIS elements in the presence of a jammer with $M=4, 8, 16$ antennas.}
    \label{fig6}
\end{figure}
In Fig.~\ref{fig6}, we show the resulting average absorption level of the A-RIS as a function of the number of A-RIS elements in the presence of a jammer with $M = 4, 8, 16$ antennas. As can be seen, the absorption capability of the A-RIS is very useful when the degrees of freedom required for the A-RIS to mitigate the jammer is low, i.e. when the number of A-RIS elements $N$ is small compared to the number of jammer antennas $M$. With increasing $N$, the average absorption level of the A-RIS elements is monotonically decreasing, and thus the operation of the A-RIS approaches that of the regular non-absorbing RIS, which is in line with the results in Figs.~\ref{fig3}-\ref{fig5} showing that the performance of a conventional RIS is approaching the performance of the A-RIS as the number of elements increases.

\begin{figure}
    \centering
    \includegraphics[height=7cm]{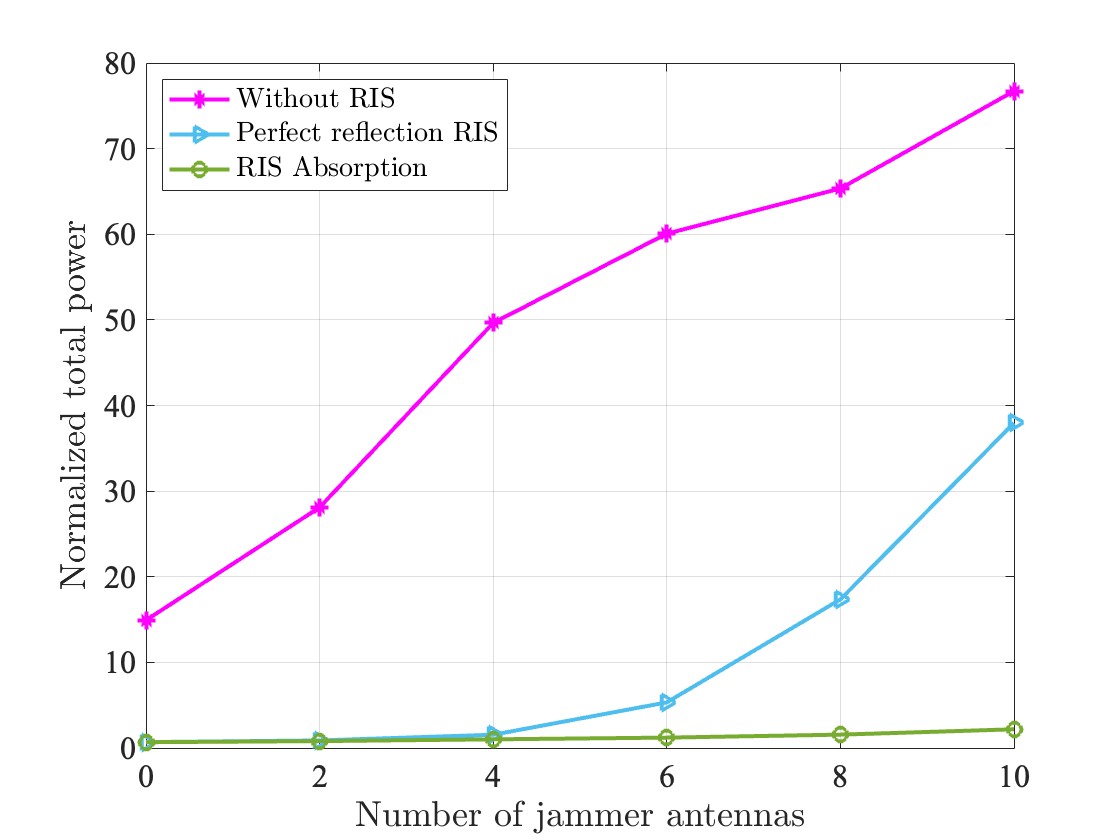}
    \caption{Required total transmit power to meet the user quality of service requirements, as a function of the number of jammer antennas, for the cases of no RIS, a regular RIS, and an A-RIS, both with $N=16$ elements.}
    \label{fig7}
\end{figure}
Finally, in Fig.~\ref{fig7}, we show the total transmit power required to meet the user quality of service requirements as a function of the number of jammer antennas for the cases of no RIS, a conventional RIS, and an A-RIS, both with $N = 16$ elements. As seen, the greater the number of jammer antennas, the more beneficial is the use of RIS technology, and the additional benefits of RIS absorption rapidly increase when the number of jammer antennas is more than $M=4$.

\section{Conclusion}\label{sec:Conclusions}
In this paper, we considered an uplink power-domain NOMA system assisted by an RIS in the presence of a smart jammer. We considered the performance of both a conventional RIS whose elements adjust only the phase of the incident signals, as well as an A-RIS that can adaptively adjust both the modulus and the phase of the elements. An optimization problem was formulated in which the goal is to minimize the total transmit power of the users under constraints on the SINR that the users achieve at the base station. Our simulation results demonstrate that both the conventional RIS and A-RIS provide dramatic reductions in required total transmit power due to their ability to enhance the signals of interest and cancel the effects of the interference. We also demonstrated that if the number of elements in the intelligent surface is small, the A-RIS can provide additional degrees of freedom that enable increased signal enhancement and interference mitigation.

\section*{Acknowledgment}
This work has been supported by the project SEMANTIC, funded from EU's Horizon 2020 research and innovation programme under the Marie Skłodowska-Curie grant agreement No 861165, as well as by the {U.S.} National Science Foundation under grants CNS-2107182 and ECCS-2030029.

\end{document}